\documentclass[%
reprint,
superscriptaddress,
 amsmath,
amssymb,
aps,
prl,
numerical,
]
{revtex4-1}

\usepackage[USenglish]{babel} 
\usepackage [table]{xcolor}

\usepackage{graphicx} 

\usepackage{amsmath}
\usepackage{amsthm}
\usepackage{amsfonts}

\usepackage{float}
\usepackage{textcomp}
\usepackage{graphicx}
\usepackage{dcolumn}
\usepackage{bm}
\usepackage{hyperref}
\usepackage[mathlines]{lineno}

\begin{document}

\newcommand*\mycommand[1]{\texttt{\emph{#1}}}

\author{C. J. Chua}
\email{cjc204@cam.ac.uk}
\affiliation{Cavendish Laboratory, 19 JJ Thomson Avenue, Cambridge CB3 0HE, United Kingdom}
\author{M. R. Connolly}
\affiliation{Cavendish Laboratory, 19 JJ Thomson Avenue, Cambridge CB3 0HE, United Kingdom}
\affiliation{National Physical Laboratory, Hampton Road, Teddington, TW11 0LW, United Kingdom}
\author{A. Lartsev}
\affiliation{Department of Microtechnology and Nanoscience, Chalmers University of Technology, S-412 96 G\"{o}teborg, Sweden}
\author{T. Yager}
\affiliation{Department of Microtechnology and Nanoscience, Chalmers University of Technology, S-412 96 G\"{o}teborg, Sweden}
\author{S. Lara-Avila}
\affiliation{Department of Microtechnology and Nanoscience, Chalmers University of Technology, S-412 96 G\"{o}teborg, Sweden}
\author{S. Kubatkin}
\affiliation{Department of Microtechnology and Nanoscience, Chalmers University of Technology, S-412 96 G\"{o}teborg, Sweden}
\author{S. Kopylov}
\affiliation{Physics Department, Lancaster University, Lancaster LA1 4YB, United Kingdom}
\author{V. I. Fal'ko}
\affiliation{Physics Department, Lancaster University, Lancaster LA1 4YB, United Kingdom}
\author{R. Yakimova}
\affiliation{Department of Physics, Chemistry and Biology (IFM), Link\"{o}ping University, S-581 83 Link\"{o}ping, Sweden}
\author{R. Pearce}
\affiliation{National Physical Laboratory, Hampton Road, Teddington, TW11 0LW, United Kingdom}
\author{T. J. B. M. Janssen}
\affiliation{National Physical Laboratory, Hampton Road, Teddington, TW11 0LW, United Kingdom}
\author{A. Ya. Tzalenchuk}
\affiliation{National Physical Laboratory, Hampton Road, Teddington, TW11 0LW, United Kingdom}
\affiliation{Royal Holloway, University of London, Egham, TW20 0EX, United Kingdom}
\author{C. G. Smith}
\affiliation{Cavendish Laboratory, 19 JJ Thomson Avenue, Cambridge CB3 0HE, United Kingdom}

\title{Quantum Hall Effect and Quantum Point Contact in Bilayer-Patched Epitaxial Graphene}

\date{\today}

\begin{abstract}

We study an epitaxial graphene monolayer with bilayer inclusions via magnetotransport measurements and scanning gate microscopy at low temperatures. We find that bilayer inclusions can be metallic or insulating depending on the initial and gated carrier density. The metallic bilayers act as equipotential shorts for edge currents, while closely spaced insulating bilayers guide the flow of electrons in the monolayer constriction, which was locally gated using a scanning gate probe.

\end{abstract}

\maketitle

Epitaxial graphene sublimated on Si-terminated silicon carbide surface \cite{emt_09, vir_08} offers one of several possible routes towards production of scalable graphene-based devices \cite{nov_12}. This material was also shown \cite{tza_10, jan_13} to be suitable for the applications of graphene in quantum resistance metrology based upon the quantum Hall effect (QHE) - phenomenon which consists of the quantization of the Hall resistance in two-dimensional (2D) electron systems accompanied by the vanishing of the Joule dissipation \cite{kli_80}. In particular, epitaxial graphene on SiC has been used to verify the universality of the QHE by a direct comparison between the quantized resistance values in graphene and GaAs/AlGaAs \cite{jan_11}, which was proven to be accurate to a level better than 0.1 ppb \cite{jan_11p}. Precision measurements, necessary for practical application of graphene in metrology, require large-area samples sustaining a high non-dissipative current.

The main challenge here is to produce entirely single-phase epitaxial graphene on a wafer scale, without multilayer graphene inclusions, since the resistance quantization and formation of dissipation-less transport in monolayer and bilayer graphene differ from each other \cite{nov_05,zha_05,nov_06}. Bilayer inclusions in an otherwise monolayer graphene sample often nucleate along the step edges on the surface of SiC, forming stripes or isolated islands. In this paper we argue that, depending on the doping and gating of the monolayer-bilayer composite, the bilayer patches can act either as metallic shortcuts (in material doped by the SiC substrate surface to a high carrier density) or insulating islands (in low carrier density material). In the latter case, pairs of closely placed insulating bilayer inclusions can create naturally defined constrictions and point contacts in conducting monolayer graphene, and, below we study how electrical transport through such constrictions can be controlled using local electrostatic fields applied from a conducting AFM tip.

Our expectations about the QHE performance of a bilayer-patched monolayer is based upon the charge transfer model for graphene on SiC. The charge transfer between graphene and donor states on the surface of SiC has been described by the balance equation \cite{kop_10}
\begin{equation}
\gamma [A-\frac{e^2 d}{\epsilon_0}(n+n_g)-\varepsilon_F] = n+n_g,
\label{one}
\end{equation}
where $A\approx 1$~eV is the work function difference between graphene and donor states, $d\approx 2$~\AA, $n_g = CV_g/e$. Equation \ref{one} allows us to relate the monolayer carrier concentration $n_0$ determined by doping, to the density of surface donor states $\gamma$, and, then, for each given $n_0$ to relate the carrier density in monolayer and bilayer graphene to the gate voltage. For a strong magnetic field, the electron spectrum in graphene is discrete, and one should distinguish between two possible situations for the charge transfer: (a) The electron Fermi level can be pinned to one of the Landau levels (LL) $\varepsilon^{(1 \mid 2)}_N$ in the graphene spectrum: $\varepsilon^{(1)}_N=\pm\hbar v \sqrt{2N}/\lambda_B$ in the monolayer \cite{mcc_56} and $\varepsilon^{(2)}_N=\pm\hbar \sqrt{N(N-1)}/m\lambda_B^2$ in the bilayer \cite {mccf_06}, where $\lambda_B=\sqrt{\hbar/eB}$ and $v\approx 10^8$ cm/s is the 'Dirac' velocity of electrons. In such situations, graphene would be metallic or display a very small QHE breakdown current. (b) The electron Fermi level is at some energy between graphene LLs, somewhere within the band of surface states on the SiC substrate surface. In the latter situation, the filling factor in graphene is pinned at the values $\nu_1=4N+2$ for monolayers and at $\nu_2=4N$ for the bilayers, forming perfect conditions for a high-breakdown-current QHE when $\nu_1=\pm 2$. Additionally, a transverse electric field created by a combination of donor charges under the bilayer and from the top gate can open a gap in its spectrum \cite{mccf_06, mcc_06}, providing a region of insulating behaviour in the bilayer.

\begin{figure}[h]
\centering
\includegraphics[width=8.46cm,keepaspectratio]{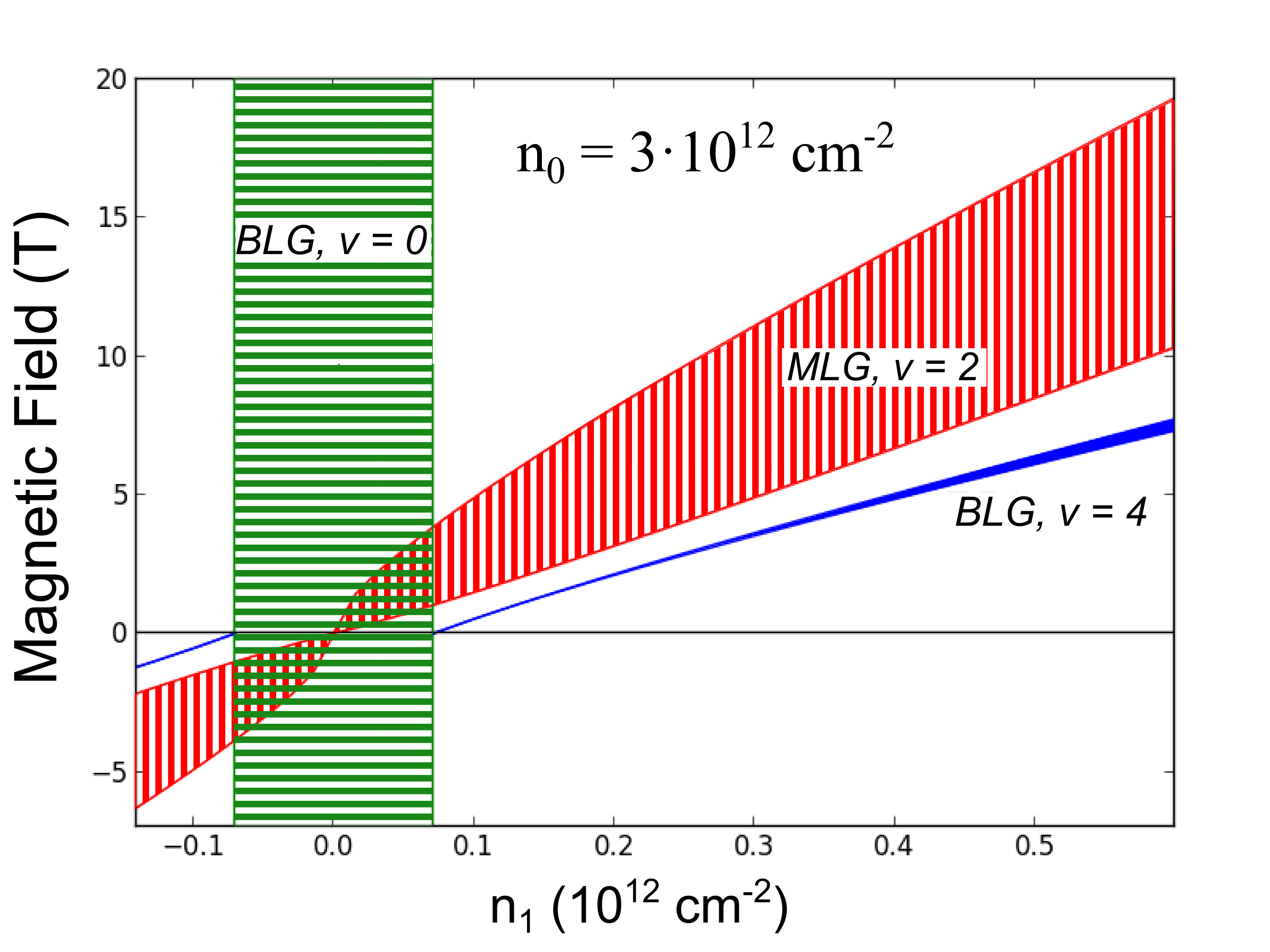}
\caption{Results obtained from electrostatic model of charge transfer from SiC substrate to monolayer and bilayer graphene for a sample with initial ungated monolayer carrier concentration of $n_0$=$3\times 10^{12}$ cm$^{-2}$. Regions of fill factor pinning in magnetic field for both monolayer graphene ($\nu$=2, red hatched region) and bilayer graphene ($\nu$=4, blue solid region and $\nu$=0, green hatched region) have been plotted as a function of gated concentration $n_1$.}
\label{fig:model}
\end{figure}

Figure \ref{fig:model} shows the calculated expectation for the appearance of type (a) and (b) conditions in monolayer and bilayer graphene on SiC with such a surface density of states of donors that dopes monolayer graphene to the density of $3 \times 10^{12}$ cm$^{-2}$, and we also take into account an electrostatic top gate voltage that reduces this density to the value $n_1$. It shows that perfect QHE conditions in monolayer and bilayer graphenes on the same SiC substrate never coincide, and in most cases, bilayers act as normal metal shortcuts for $\nu_1=\pm2$ QHE in the monolayer. The exception is only in a low-density interval (green hatched region in Fig. \ref{fig:model}) where the $\nu_1=\pm2$ QHE in the monolayer (red hatched region in Fig. \ref{fig:model}) coexists with the insulating behaviour of the bilayer prescribed by the interlayer asymmetry gap opened by the transverse electric field in its spectrum \cite{mccf_06}. In this case, bilayer inclusions would act as borders of the monolayer channel without destroying the quantization of the Hall resistance in it, unless they form a continuous cluster cutting across parts of the Hall-bar device.

\begin{figure}[h]
\centering
\includegraphics[width=8.46cm,keepaspectratio]{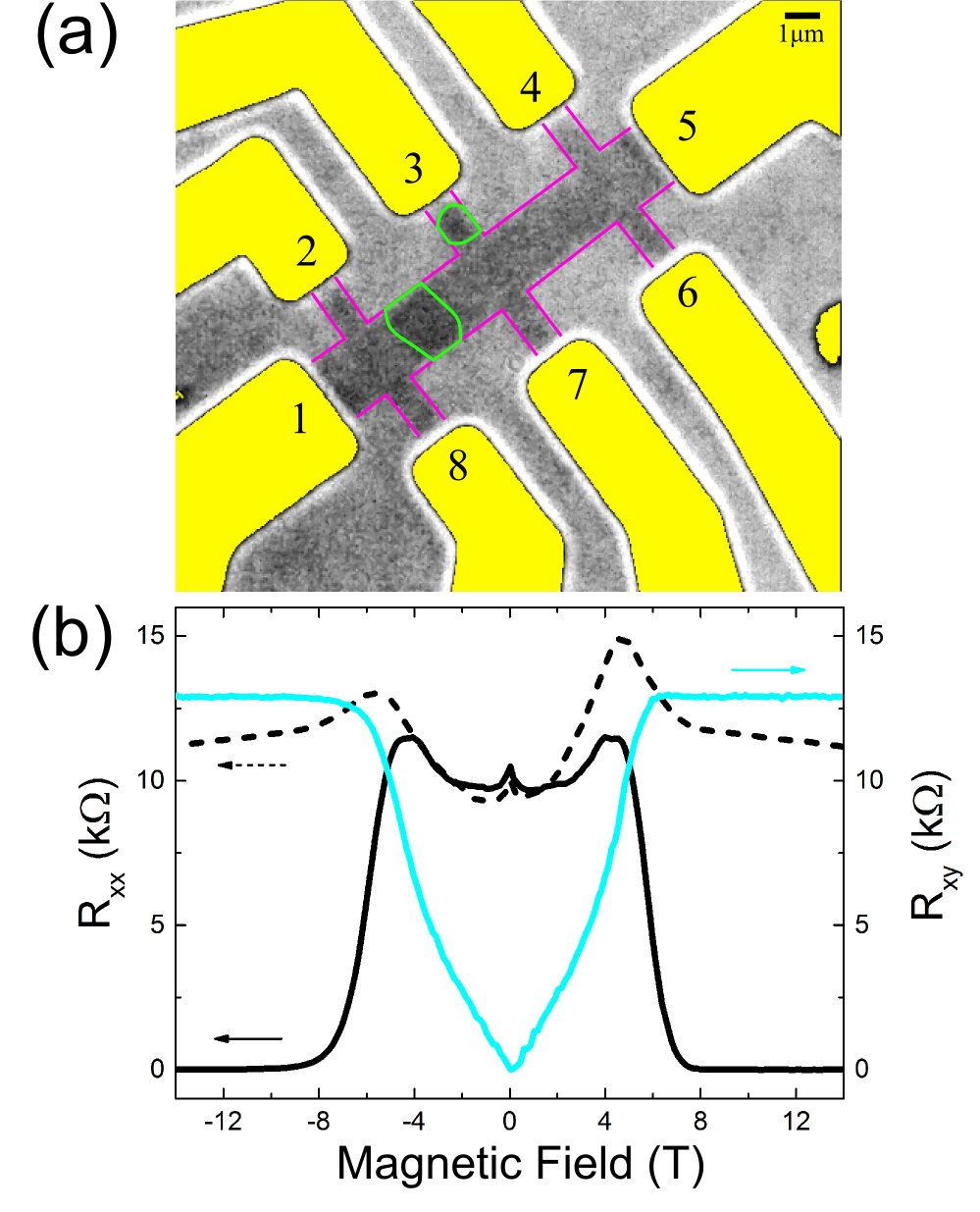}
\caption{(a) An optical transmission micrograph of device 1 with photo-gated carrier concentration of $n_1$ = $~10^{11}$ cm$^{-2}$ with graphically superimposed device layout; dark gray regions are bilayers (outlined in light green) and light gray regions are monolayers, with a natural contrast of about ~1.3\%. (b) the corresponding transverse resistance (light blue) and longitudinal resistance (black) plots measured from contacts 4-6 and 6-7, respectively, in the entirely monolayer region at T=4.2 K. The black, dashed plot is the (nominally) longitudinal resistance measured from contacts 8-7 separated by a bilayer patch.}
\label{fig:metallic}
\end{figure}

The magneto-transport experiments reported in this work confirm the model described above. These experiments were performed on several samples with high and low densities, where bilayer patches either cross the sample in a single well-defined place or flank the monolayer channel creating a narrow constriction in it. These samples have been grown by high-temperature sublimation on the Si face of semi-insulating SiC. One sample contained predominantly monolayer graphene with a low density of bilayer patches along the substrate vicinal steps, and the Hall bar device, with a narrow monolayer stripe oriented perpendicular to the vicinal edge steps, was fabricated in such way that only a single bilayer patch crossed the bar. An optical micrograph \cite{yag_13} of the first sample with elongated bilayer patches clearly visible is shown in figure~\ref{fig:metallic}a. The second sample contained a large number of patches and was fabricated along the steps. The carrier density in both samples, initially close to $n_0 = 3 \times 10^{12}$ cm$^{-2}$, was reduced by photochemical gating \cite{lar_11} to $n_1$ = $~10^{11}$ cm$^{-2}$ and approximately $5 \times 10^{10}$ cm$^{-2}$ respectively.

When measured on the entirely monolayer half of the first sample the magneto-transport plots in figure~\ref{fig:metallic}b show quantized Hall resistance above about 7.5 T (light blue curve) and the co-incidental vanishing of the longitudinal resistance (black curve). However, when measured across the bilayer patch, the longitudinal resistance (dashed black curve) rather than vanishing, saturates at the level close to $h/2e^2$. This behavior is consistent with shunting of the edge channels by the metallic bilayer patch \cite{lof_13}, as predicted by our electrostatic model for this high carrier concentration in the gated device.

\begin{figure}[h]
\centering
\includegraphics[width=8.46cm,keepaspectratio]{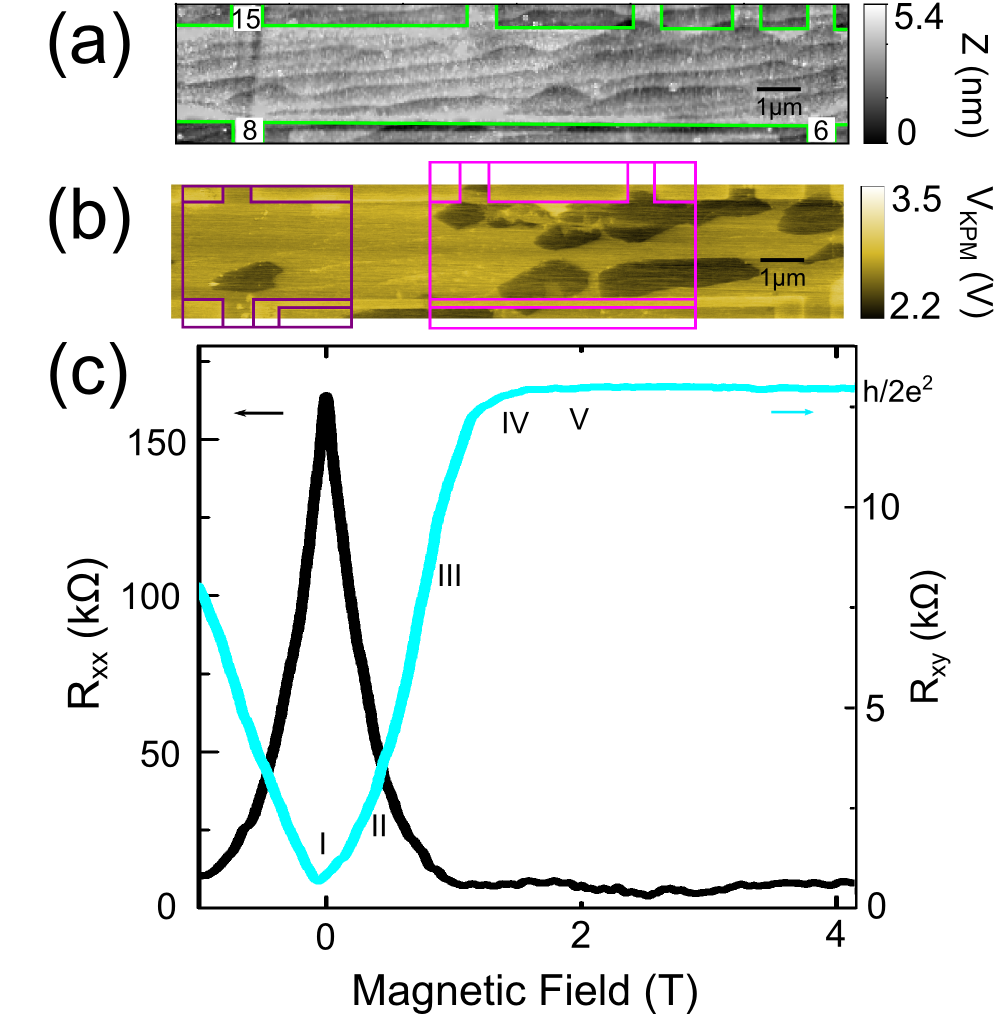}
\caption{(a) AFM amplitude image of device 2 with photo-gated carrier concentration of $n_1$ = $5\times 10^{10}$ cm$^{-2}$. (b) SKPM image of the device, showing regions of monolayer (light gray) and bilayer (dark gray) graphene. Scan regions are outlined in purple (left region) and pink (middle region). (c) Magnetoresistance plot of the device measured at T=4.2 K, with transverse resistance (light blue) measured from contacts 15-8 and longitudinal resistance (black) measured from contacts 6-8. Regions scanned using SGM are also outlined and colour-coded with dark purple for left side region, and light purple for middle region.}
\label{fig:magnetoresistance}
\end{figure}

The geometry and topography of the second sample, visualized by atomic force microscopy (AFM) and room-temperature scanning Kelvin probe microscopy \cite{bur_11} (SKPM), are shown in figure~\ref{fig:magnetoresistance}. SKPM enables us to identify monolayer (light shade) and bilayer (dark shade) regions in the device. The bilayer patches cluster in the right-hand side of the device, leaving a narrow monolayer channel in between.

Magneto-transport measured at a temperature of 4.2 K in the second sample shows a (nearly) quantized Hall resistance and the longitudinal resistance which drops by a factor of 40 from an extremely large zero-field value of 160 k$\Omega$ compared to sample 1. The $\nu$ = 2 plateau is reached at approximately $\pm$ 1 T, due to the low carrier concentration, estimated as $n_1 = 5 \times 10^{10}$ cm$^{-2}$. At this low carrier density the result of our model shown in fig. \ref{fig:model} suggests that the bilayer patches should be in the gapped insulating state, so that the electronic transport occurs through a network of narrow monolayer channels squeezed between insulating bilayer patches - in agreement with the measured high resistance of the sample.

In quantizing magnetic fields, the current through the narrow constrictions squeezed between insulating bilayer patches is carried by the edge channels with a strongly reduced backscattering \cite{mol_09}. We note that at $B=1$ T, magnetic length $\lambda_B =25$ nm is already much less than the width of the narrowest constriction, $w \approx 200$ nm for the sample in question (see Fig.\ref{fig:magnetoresistance}), and the edge channels are well separated. This condition corresponds to the formation of the $\nu=2$ QHE plateau shown in fig. \ref{fig:magnetoresistance}, which then extends over the interval of several Tesla, as it is protected by the charge transfer from the states localised at the substrate surface and the resulting $\nu =2$ filling factor pinning in monolayer graphene \cite{jan_11p}.

\begin{figure}[H]
\centering
\includegraphics[width=8.46cm,keepaspectratio]{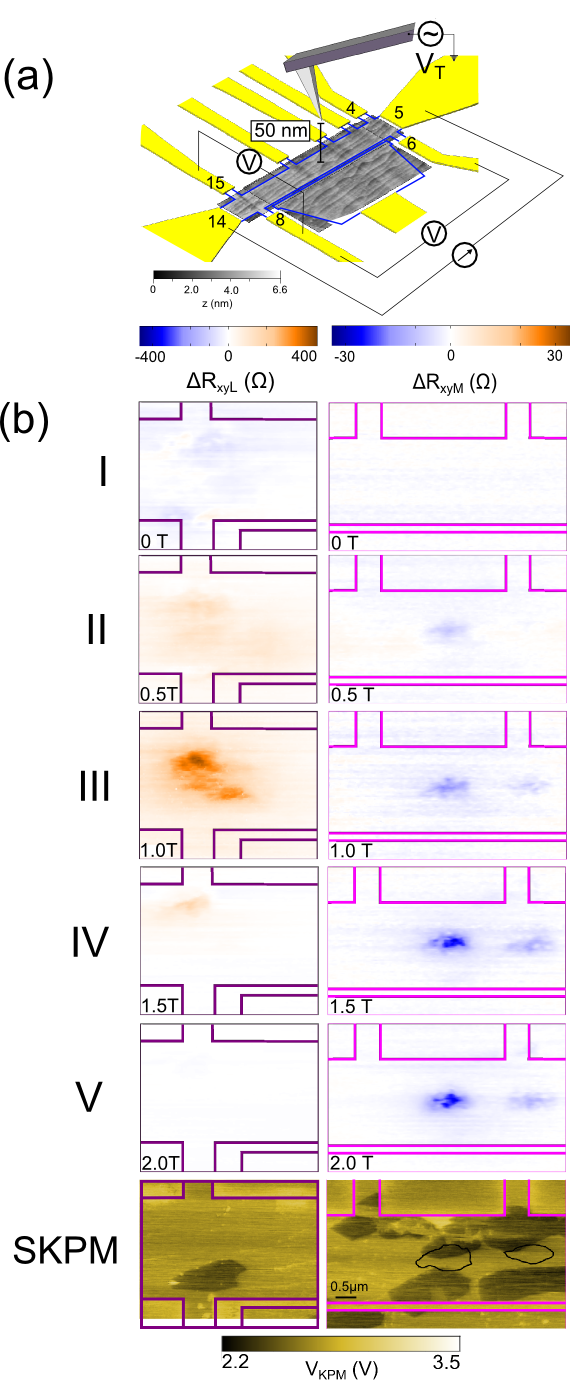}
\caption{(a) Schematic of SGM measurement set-up used for device 2. (b) Sequence of SGM images taken in different magnetic fields indicated by roman numerals in fig. \ref{fig:magnetoresistance} and measured from contacts 15-8 while scanning the left region of device (left column); measured from contacts 15-8 while scanning middle region of device (right column) at T=4.2 K (see figure\ref{fig:magnetoresistance}b for scan regions). The last image in each sequence is an SKPM image of the corresponding region with the area of the strongest SGM response outlined.}
\label{fig:sgm}
\end{figure}

Based on the latter observation, we seize the opportunity to explore the properties of a new type of point contact (narrow channel) in monolayer graphene formed where it is flanked by bilayer inclusions (see figure~\ref{fig:magnetoresistance}b). For this, we use low temperature scanning gate microscopy (SGM). In brief, SGM (see figure~\ref{fig:sgm}a) involves scanning a sharp metallic tip over the surface of graphene while measuring its conductance. A schematic of our SGM setup is shown in fig.\ref{fig:sgm}a. The oscillation of the cantilever is measured using standard interferometric detection with a fibre-based infra-red laser. We use a Pt/Ir coated cantilever with a nominal tip radius of 15 nm. In order to avoid any cross-contamination between the tip and the sample during SGM, once the sample is approached using tapping mode, we switch to lift mode with the static tip at a lift height of approximately 50 nm.

Scanning gate images were taken from two main regions of sample 2: the left hand side region, where there is mostly monolayer graphene, and in the middle region of the device where bilayer graphene is predominant. The left column of SGM images (see figure~\ref{fig:sgm}b) were measured from transverse contacts 15-8 while scanning the left-hand side region of the device with an AC tip voltage of 2 V at various values of magnetic field. We find that the strongest SGM signal occurs at around 0.8 T, on the riser before the $\nu =2$ plateau, and it gradually vanishes deeper on the plateau. This is consistent with the percolation of single-particle orbits in non-interacting electron picture of the quantum Hall breakdown\cite{con_12} (see supplementary information). On the other hand, for scans performed at the middle region of the device where monolayer graphene is closely flanked by large bilayer inserts (figure~\ref{fig:sgm}b, right column), the response increases with magnetic field until the plateau is reached and stays nearly independent of the magnetic field throughout the plateau.

\begin{figure}[H]
\centering
\includegraphics[width=8.46cm,keepaspectratio]{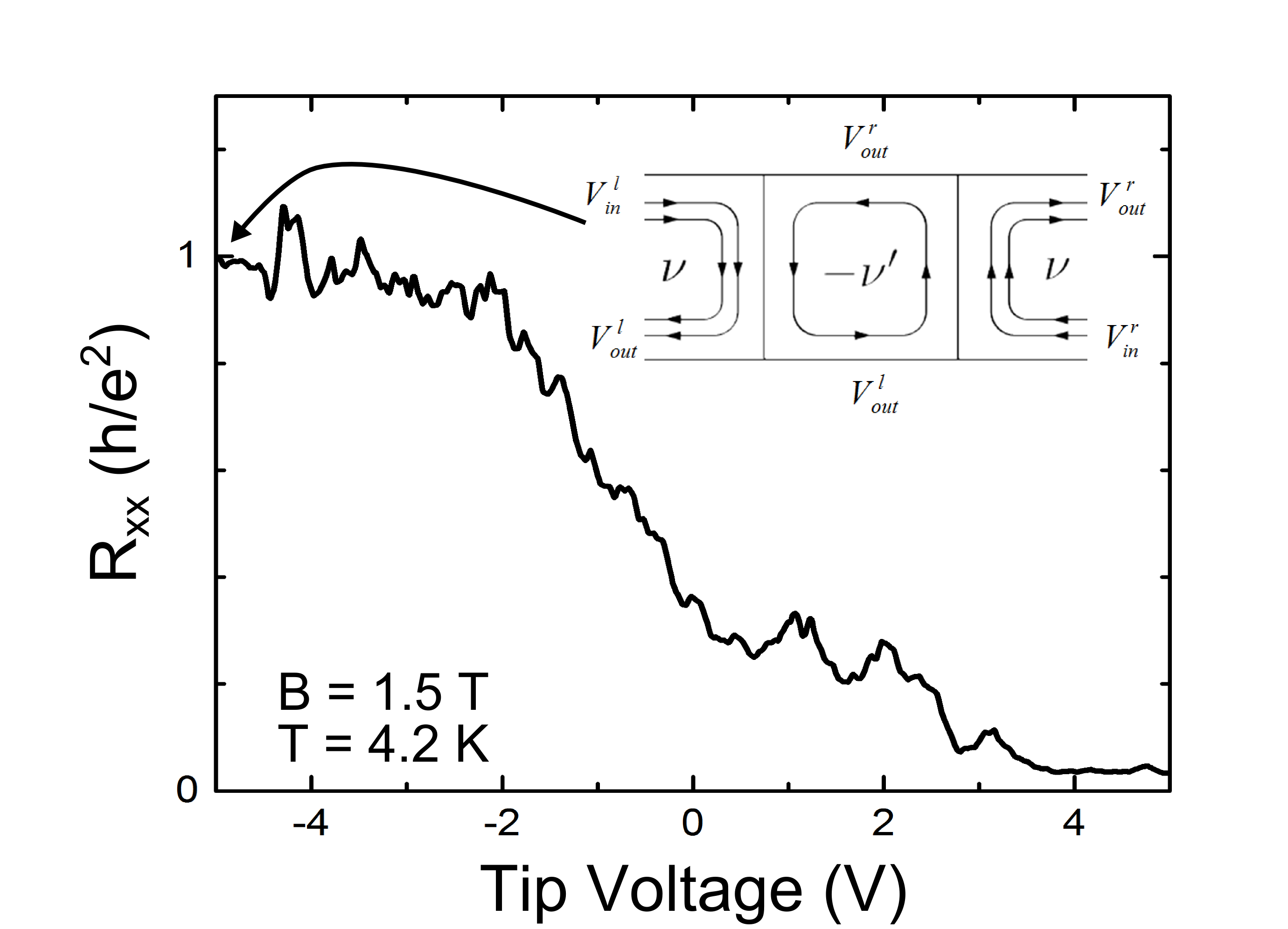}
\caption{Longitudinal resistance measured from contacts 6-8 as a function of tip voltage when sat above the monolayer constriction in device 2. Inset: Electron edge state propogation in an n-p-n junction}
\label{fig:npn}
\end{figure}

To understand this behavior better, we studied the dependence of the longitudinal resistance $R_{6-8}=(V_{6}-V_{8})/I$ between two contacts 6 and 8 closest to the bilayer-confined constriction in the monolayer on the magnitude of the applied scanning gate voltage locally changing the carrier density in the constriction independently of the rest of the sample. Figure~\ref{fig:npn} shows the variation in the longitudinal resistance  $R_{6-8}$ at $B=1.5$ T (where the device already shows the QHE behavior, fig. \ref{fig:magnetoresistance}c) as a function of tip bias as the tip is sat above the monolayer constriction. The value of $R_{6-8}$ varies between nearly zero, corresponding to the non-dissipative quantum Hall effect transport at a positive tip bias to almost exactly $h/e^2$ at a large negative tip potential.

The increase of resistance  $R_{6-8}$ indicates that the local gate has introduced scattering of electrons between the edge channels propagating along bilayer boundaries of the constriction, whereas its saturation at the $h/e^2$ value at a high negative tip voltage can be explained if we assume that the point contact in graphene has been driven into an n-p-n bipolar state, with the density reaching filling factor $\nu =-2$ in the middle. The value $h/e^2$ is the result of the equilibration of electron edge states propagating along p-n junctions in an n-type doped QHE wire with the filling factor $\nu$, cut across by a p-doped region with filling factor $-\nu'$ \cite{ozy_07, wil_07, ki_09, nak_11}. The equivalent electrical scheme describing chemical potentials at different parts of the edges of such a QHE wire is sketched in the inset in fig. \ref{fig:npn}, where the incoming channels at voltages $V_{in}^r$ and $V_{out}^l$ would correspond to the measured voltages $V_8$ and $V_{15}$, respectively, as used to define the resistance $R_{8-15}$ in the Hall bar shown in fig. \ref{fig:magnetoresistance}a. The continuity of electric current in this equivalent electrical scheme dictates that
\begin{equation}
I = \frac{\nu e^2}{h}(V_{in}^l - V_{out}^l) = \frac{\nu' e^2}{h}(V_{out}^l - V_{out}^r) = \frac{\nu e^2}{h}(V_{out}^r - V_{in}^r), \nonumber
\end{equation}
where we used the quantization of Hall conductivity in each part of the n-p-n QHE circuit, and assumed a complete equilibration between edge currents arriving at each of the two p-n junctions and propagating along it between the opposite edges (that is how chemical potentials $V_{out}^r$ and $V_{out}^l$ appear at the edges of the p-doped region in the middle of this QHE wire). As a result, we find
\begin{equation}
R_{6-8} \equiv (V_{out}^l - V_{in}^r)/I = \frac{\nu+\nu'}{\nu\nu'} \frac{h}{e^2} \xrightarrow[\nu=\nu'=2]{} \frac{h}{e^2},
\end{equation}
as shown in figure~\ref{fig:npn}.

In conclusion, we have demonstrated the important role that bilayer inclusions in epitaxial monolayer graphene play in forming its magneto-transport characteristics. The effect of bilayer inclusions depends on their position within the device, and differs for low and high carrier density samples. For a high carrier concentration, bilayer inclusions behave as metallic shunts for the quantum Hall edge states, if the inclusions run across the device perpendicular to the direction of Hall current. For low-density samples, where the carrier density is strongly reduced by the top gate, the bilayer inclusions are driven to the insulating state due to the interlayer asymmetry gap promoted by the transverse electric field.  This permits the dissipation-less QHE transport to survive in the samples where the bilayer inclusions do not cut off the contacts. Moreover, pairs of large-area insulating bilayer islands can be used to produce new types of nano-scale devices in monolayer graphene.

We acknowledge support form European Commission ICT STREP ConceptGraphene, FET Graphene Flagship, ERC Synergy grant Hetero2D, EMRP GraphOhm, Swedish SSF, Knut and Alice Wallenberg Foundation, Linneqs Center for Engineered Quantum Systems, Chalmers Areas of Advance, and EPSRC.

\bibliography{Cassandra}

\end{document}